\documentclass[twocolumn,preprintnumbers,prl]{revtex4}
\usepackage{amsmath,amssymb,bm,graphicx,epsfig,color}
\usepackage{feynmf}
\usepackage{psfrag}
\usepackage{graphicx}
\usepackage{bm}

\newcommand{\be}{\begin{equation}}  
\newcommand{\ee}{\end{equation}} 
  
\newcommand{\nl}{\nonumber \\ }

\newcommand{\order}{{\cal O}}

\begin{document}

\preprint{EFI Preprint 11-9}

\title{Model independent analysis of proton structure for hydrogenic bound states}

\author{Richard J. Hill and Gil Paz}

\affiliation{Enrico Fermi Institute and Department of Physics \\
The University of Chicago, Chicago, Illinois, 60637, USA}

\date{March 23, 2011}

\begin{abstract}
Proton structure effects in hydrogenic bound states are
analyzed using  nonrelativistic QED effective field theory.
Implications for the Lamb shift in muonic hydrogen are
discussed. 
Model-dependent assumptions in previous analyses are isolated, and 
sensitivity to poorly constrained hadronic structure  
in the two-photon exchange contribution is identified. 
\end{abstract}

\vspace{-5mm}
\maketitle

\allowdisplaybreaks

\vspace{-5mm}
\noindent
{\bf Introduction.}
Atomic spectroscopy can provide the most precise determination of
fundamental hadron properties, such as the proton 
radius~\cite{Udem:1997zz,Pohl:2010zz,Mohr:2008fa}.    
The need for systematic analysis 
to translate between bound state energies and hadronic observables
is sharpened by a 
discrepancy between the recent
muonic hydrogen Lamb shift measurement~\cite{Pohl:2010zz}
and existing theoretical calculations.
Using a model-independent extraction of the
charge radius from electron scattering data 
($r=0.871(10)$~fm \cite{Hill:2010yb}; see also \cite{Bernauer:2010wm,Zhan:2011ji} ) 
or an extraction from electronic 
hydrogen spectroscopy ($r=0.8768(69)$~fm \cite{Mohr:2008fa}; see also \cite{CODATA2010}), the measured 
$2S_{j=1/2}^{(F=1)}-2P_{j=3/2}^{(F=2)}$ interval in muonic hydrogen lies 
$0.258(90)\,{\rm meV}$ or $0.311(63)\,{\rm meV}$ above theory. 
The discrepancy brings into
question the treatment of proton structure effects in  atomic bound
states, and has generated speculations on new forces acting in the
muon-proton system~\cite{Barger:2010aj},
inadequate treatment of proton charge density correlations~\cite{DeRujula:2010dp},
and modifications of offshell photon vertices~\cite{Miller:2011yw}.

Non-relativistic QED (NRQED)~\cite{Caswell:1985ui}  is a field theory
describing the interactions of  photons and nonrelativistic
matter.  
The NRQED lagrangian  is constructed to yield predictions
valid to any fixed order in  small parameters $\alpha$ and
${|\bm{q}|/M}$, where $|\bm{q}|$ denotes a typical  bound state
momentum, and $M$ is a mass scale for the nonrelativistic particle.
NRQED provides a
rigorous framework to study the effects of  proton structure, avoiding
problems of double counting in bound  state energy
computations~\cite{Jentschura:2010ty}; eliminating difficulties of
interpretation for  the polarizability of a strongly interacting
particle~\cite{Lvov:1993fp};  and providing trivial derivations of
universal properties,  such as the low energy theorems of Compton
scattering~\cite{Ragusa:1993rm}.  

We examine the NRQED
framework for determining proton structure corrections in atomic bound
states.  
The Lamb shift in muonic hydrogen is the first measurement directly sensitive to 
the spin-independent, proton structure-dependent, contact interaction 
appearing in NRQED ($d_2$ below).    
The strength of this interaction is not determined by measured on-shell 
form factors, or inelastic structure functions of the proton. 
We identify model-dependent assumptions in previous analyses and  discuss whether
poorly constrained proton structure corrections can account for the
above-mentioned discrepancy.  We conclude by outlining
extensions of the theoretical analysis and related applications. 

\noindent 
{\bf NRQED.}
Consider the formalism for 
electron-proton bound states; the substitution $e\to \mu$ applies 
for the muon-proton
system.  
The NRQED lagrangian 
can be decomposed as 
${\cal L}_{\rm NRQED} = {\cal L}_\gamma + {\cal L}_e +  {\cal L}_p + {\cal L}_{\rm contact}$.
${\cal L}_\gamma$ contains the photon kinetic term
and   vacuum  polarization  corrections; 
these corrections can be treated separately and 
will not be considered here.    
Through $\order(1/m_e^3)$,~\cite{Caswell:1985ui,Kinoshita:1995mt,Manohar:1997qy}
\begin{widetext}
\vspace{-8mm}
\begin{multline} 
\label{eq:NRQED} 
  {\cal L}_{e} = \psi_e^\dagger
  \bigg\{  i D_t  + {\bm{D}^2 \over 2 m_e}  + {\bm{D}^4 \over 8 m_e^3} +
  c_F e{ \bm{\sigma}\cdot \bm{B} \over 2m_e}   
+ c_D e{ [\bm{\partial}\cdot \bm{E} ]  \over 8 m_e^2}  + i c_S e{ \bm{\sigma}
    \cdot ( \bm{D} \times \bm{E} - \bm{E}\times \bm{D} ) \over 8 m_e^2} 
  + c_{W1}e {  \{ \bm{D}^2 ,  \bm{\sigma}\cdot \bm{B} \}  \over 8 m_e^3}  
\\
- c_{W2}e {  D^i \bm{\sigma}\cdot
    \bm{B} D^i \over 4 m_e^3 }  + c_{p^\prime p} e { \bm{\sigma} \cdot
    \bm{D} \bm{B}\cdot \bm{D} + \bm{D}\cdot\bm{B} \bm{\sigma}\cdot \bm{D}
    \over  8 m_e^3} 
+ i c_M e { \{ {D}^i ,  [\bm{\partial} \times
    \bm{B}]^i \}
\over 8 m_e^3}
 +
  c_{A1} e^2{ \bm{B}^2 - \bm{E}^2 \over 8 m_e^3}  - c_{A2} e^2{ \bm{E}^2
    \over 16 m_e^3 } 
 \bigg\} \psi_e  \,.
\end{multline} 
\end{widetext}
Here $\psi_e$ is a two-component spinor representing the
nonrelativistic electron field, $\bm{\sigma}$ is the Pauli spin matrix,
$D_t$ and $\bm{D}$ are covariant derivatives
and $\bm{E}$, $\bm{B}$ are the electric and magnetic fields.
Prefactors are chosen for convenience so that
for a point-like fermion at tree level, $c_F=c_D=c_S=c_{W1}=c_{A1}=1$
and $c_{W2}=c_{p^\prime p}=c_M=c_{A2}=0$.  A similar expression holds
for ${\cal L}_p$
with $e\to -Ze$ ($Z=1$ for the proton).     
Relevant contact interactions in the single proton plus single
electron  sector are
\begin{align}\label{eq:contact} {\cal L}_{\rm contact} &= d_1
{\psi_p^\dagger \bm{\sigma} \psi_p \cdot  \psi_e^\dagger \bm{\sigma} \psi_e \over m_e m_p} 
+ d_2 {\psi_p^\dagger \psi_p
\psi_e^\dagger  \psi_e \over m_e m_p} \,. 
\end{align} 
The coefficients $c_i$, $d_i$ 
depend on  the choice of ultraviolet regulator.    
Since no new bound state computations are necessary, 
we will quote results for phenomenological inputs and bound state energies 
that are independent of this choice. 
We are interested in proton
structure corrections to  energy levels through order
$m_e^3\alpha^5/m_p^2$, and therefore 
need  $c_{F,D,S}$ in ${\cal L}_p$ 
through $\order(\alpha)$, and
$d_{1,2}$ through $\order(\alpha^2)$.  Other operators 
in ${\cal L}_p$ will enter when we analyze the low $Q^2$ expansion of  
the forward Compton amplitude to constrain $d_2$.
Knowledge of the $c_i$'s and $d_i$'s allows us to determine corrections to energy levels. 
For example, coefficients $c_D^{\rm proton}$ and 
$d_2$ lead to first order energy shifts
\be\label{eq:Enl}
\delta E(n,\ell) = \delta_{\ell 0} {m_r^3 (Z\alpha)^3 \over \pi n^3} \left( 
{Z\alpha \pi \over 2 m_p^2} c_D^{\rm proton}   - {1\over m_e m_p} d_2  \right) \,,
\ee
where $m_r = m_e m_p/(m_e+m_p)$ is the reduced mass.

\noindent
{\bf Matching.}
The NRQED Wilson coefficients are determined 
by enforcing matching conditions between full and effective theories 
using convenient low energy observables. 
We concentrate on the matching conditions for the proton. 

\noindent
\emph{One photon matching.}
Wilson coefficients for operators coupling to a single photon
are determined 
in terms of the proton elastic form factors and their derivatives
at $q^2=0$
by using (\ref{eq:NRQED}) to compute  the amplitude for
elastic scattering of a proton via the electromagnetic 
current~\cite{Manohar:1997qy, HillPaz, footnotesign}.
The form factors satisfy 
$F_1(0)= 1$, $F_2(0) = a_p$,
\begin{align}\label{eq:re_define}
F_1^\prime(0) &= 
\frac16 (r^p_E)^2 -\frac{a_p}{4m_p^2} 
+ {Z^2\alpha \over 3 \pi m_p^2} \log{m_p\over \lambda} \,,
\nl
F_2^\prime(0) &= 
\frac16 \big[ (1+a_p) (r^p_M)^2 - (r_E^p)^2 \big] + \frac{a_p}{4 m_p^2} \,,
\end{align} 
where $\lambda$ is a photon mass~\cite{footnotelambda}.
These expressions
serve to define the phenomenological parameters 
$a_p\approx 1.793$, 
$r_E^p$ and $r_M^p$.

\begin{figure}
\epsfig{file=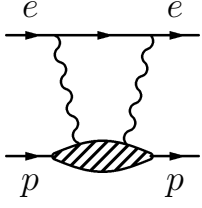,width=3.1cm}
\caption{\label{fig:blob}Two photon exchange amplitude for $e^- p \to e^- p$ scattering.}
\end{figure}

\noindent
\emph{Two photon matching.}
The coefficients $c_{A1}$, $c_{A2}$ can 
be determined by comparing to 
spin-averaged amplitudes for forward and backward Compton scattering in the lab 
frame~\cite{HillPaz}
\begin{align}
4m_p^3 \bar{\alpha}/\alpha &= -c_{A1} - c_{A2}/2 + 1 + 2 c_M + c_F c_S - c_F^2 \,, 
\nl
4m_p^3 \bar{\beta}/\alpha &= c_{A1} - 1 \,,
\end{align}
where
$\bar{\alpha}=12.0(6)\times 10^{-4}{\rm fm}^3$ 
and $\bar{\beta}=1.9(5)\times 10^{-4}{\rm fm^3}$~\cite{Nakamura:2010}.

\noindent
\emph{Contact interactions.}
The coefficients in (\ref{eq:contact}) can be fixed using the
zero-momentum limit for $e^- p \!\to\! e^- p$ scattering, cf. Fig.~\ref{fig:blob}.  
The tree level, $\order(\alpha)$, amplitude is reproduced by the effective field theory, 
and the $d_i$'s receive a non-zero contribution starting at $\order(\alpha^2)$.
We focus on the spin-independent case and neglect higher order radiative corrections.  
The relevant proton matrix element is the forward Compton amplitude 
($\nu = 2k\cdot q$, $Q^2=-q^2$)  
\vspace{-2mm}
\begin{align}\label{eq:forwardcompton}
&\frac12 \sum_s i \int d^4x\, e^{iq\cdot x} \langle \bm{k},s| T\{ J_{\rm e.m.}^\mu(x) J_{\rm e.m.}^\nu(0) \} | \bm{k},s \rangle 
\nl
&= 
\left( - g^{\mu\nu} + q^\mu q^\nu/ q^2 \right) W_1(\nu,Q^2)
\nl
&
+ 
\left( k^\mu - k\cdot q \,q^\mu /q^2 \right) 
\left( k^\nu - k\cdot q \, q^\nu /q^2 \right) W_2(\nu,Q^2) \,.
\end{align} 
Our normalizations are such that for a point particle, 
$W_1=2\nu^2/(Q^4-\nu^2)$ and $W_2=8Q^2/(Q^4-\nu^2)$.  
The matching condition for the spin-averaged zero-momentum four-point 
amplitude is
\begin{widetext}
\vspace{-7mm}
\begin{align}\label{eq:threshold}
&
{4\pi m_r \over \lambda^3} \!-\! {\pi m_r \over 2 m_e m_p\lambda}  \!-\! { 2\pi m_r \over m_p^2\lambda}\left[ F_2(0) 
\!+\! 4m_p^2 F_1^\prime(0) \right] 
- {2\over m_e m_p} \bigg[ 
\frac23 + {1\over m_p^2 - m_e^2}\left( m_e^2 \log{m_p\over \lambda} - m_p^2 \log{m_e\over\lambda} \right)
\bigg] 
+ {\delta d_2 (Z\alpha)^{-2}\over m_e m_p} 
\nl
\!&=\! -
{m_e\over m_p} \int_{-1}^1 dx   \sqrt{1-x^2} 
\int_0^\infty dQ \, {Q^3 
\left[ (1+2x^2)W_1( 2im_pQ x ,Q^2) - (1-x^2) m_p^2 W_2( 2im_pQ x, Q^2 ) \right]
\over (Q^2 + \lambda^2)^2(Q^2 + 4 m_e^2 x^2)} \,, 
\end{align}
\end{widetext}
where $\delta d_2$ denotes the contribution 
to $d_2$ in addition to the point particle value. 

The imaginary part of the $W_i$'s can be related to
measured quantities. By inserting a complete set of states into
(\ref{eq:forwardcompton}),  
the  proton contribution to ${\rm Im}\,W_i$  is 
expressed in terms of proton form factors, 
and the continuum contribution to  ${\rm Im}\, W_i$ is determined by inelastic
structure functions. 
Using dispersion relations, $W_2$ can be fully reconstructed from its 
imaginary part. Since $W_1$ requires a subtraction for a convergent dispersion relation, 
knowledge of $W_1(0,Q^2)$ is also needed.  Thus $W_i$ can be written
\begin{align}\label{eq:disp}
W_1(\nu,Q^2) &= W_1(0,Q^2) + W_1^{p, 1}(\nu,Q^2) + W_1^{c, 1}(\nu,Q^2) \,,
\nl
W_2(\nu,Q^2) &= W_2^{p, 0}(\nu,Q^2) + W_2^{c, 0}(\nu,Q^2) \,,
\end{align} 
where the superscript numbers denote the number of subtractions. 
The proton terms are 
\begin{align}\label{eq:Wiproton}
\!\!W_1^{p, 1}(\nu,Q^2) \!&= \!{2 \nu^2  (F_1 + F_2)^2/ ( Q^4 - \nu^2)},
\nl
\!\!W_2^{p, 0}(\nu,Q^2) \!&= \! 2Q^2 (4 F_1^2 
+ Q^2 F_2^2/m_p^2)/(Q^4 - \nu^2),
\end{align} 
with $F_i\equiv F_i(-Q^2)$. 
The continuum terms are
\begin{align}
W_1^{c,1}(\nu,Q^2) &= {\nu^2\over \pi} \int_{\nu_{\rm cut}(Q^2)^2}^\infty { d\nu^{\prime 2}} 
{ {\rm Im}W_1(\nu^\prime, Q^2) \over \nu^{\prime 2} (\nu^{\prime 2} - \nu^2) } \,,
\nl
W_2^{c,0}(\nu,Q^2) &={1\over \pi} \int_{\nu_{\rm cut}(Q^2)^2}^\infty { d\nu^{\prime 2} }
{ {\rm Im}W_2(\nu^\prime, Q^2) \over \nu^{\prime 2} - \nu^2} \,,
\end{align} 
where $\nu_{\rm cut}(Q^2)=Q^2 + 2 m_\pi m_p + m_\pi^2$ 
is the threshold for pion production, and ${\rm Im}W_i(\nu,Q^2)$ are proportional to 
inelastic scattering cross sections. 

The small and large $Q^2$ limits of  $W_1(0,Q^2)$ can be studied in a model-independent way.
Using NRQED to compute the amplitude for double scattering of 
a proton in an external static magnetic field we find~\cite{HillPaz}
\begin{align}\label{eq:lowQ2}
&W_1(0,Q^2) 
= 2 a_p(2 + a_p) 
+
Q^2 \big\{
 { 2m_p } {\bar{\beta}/ \alpha}
- {a_p /m_p^2} 
\nl
&
-(2/3)\left[ (1+a_p)^2 (r_M^p)^2 - (r_E^p)^2 \right] 
\big\} + \order(Q^4)\,.
\end{align} 
At $Q^2\gg {\rm GeV}^2$ 
we may evaluate 
(\ref{eq:forwardcompton}) 
using the operator product expansion (OPE).
Leading terms arise from dimension-four operators
and scale as $Q^{-2}$~\cite{HillPaz}.
The intermediate $Q^2$ region is not constrained by existing measurements.
This lack of knowledge about $W_1(0,Q^2)$ introduces model-dependence 
in the theoretical prediction for the Lamb shift, which has so far been ignored in the literature.  

Given this model-dependence, how did previous studies obtain
numerical predictions? The most common approach is to pretend that
$W_1(0,Q^2)$ can be separated into ``proton'' and ``non-proton''
contributions. A ``proton'' part for $W_1(0,Q^2)$ is obtained by
inserting the vertex with onshell form factors into the Feynman
diagrams for a relativistic pointlike particle. 
For definiteness we refer to this approach 
as the ``Sticking In Form Factors'' (SIFF) model.  Explicitly, 
\be\label{eq:SIFF} 
W_1^{\rm SIFF}(0,Q^2) = 2F_2( 2 F_1 + F_2) \,.  
\ee 
We emphasize that (\ref{eq:SIFF}) is not derived from a well
defined local field theory. In fact, \emph{no} local lagrangian can give such Feynman rules.  
Note also that $W_1^{\rm SIFF}(0,Q^2)$ does not have the correct large $Q^2$ behavior. 
A ``non-proton'' part is obtained by multiplying the 
$2m_p  Q^2 {\bar{\beta}/ \alpha}$ term in (\ref{eq:lowQ2}) 
by a function of $Q^2$~\cite{Pachucki:1999zza, Carlson:2011zd}.
The models used in \cite{Pachucki:1999zza, Carlson:2011zd} again do not have  the correct
large $Q^2$ behavior. 
Unlike ${\rm Im}\,W_1$, we stress that the separation of $W_1(0,Q^2)$ into 
proton and non-proton parts is not well-defined.

\noindent \textbf{Bound state energies.}  
The use of an effective field theory allows us to systematically classify the proton
structure corrections to energy levels. 
Using (\ref{eq:Enl}), proton vertex 
corrections, of order $(Z\alpha)^4$ and $(Z^2\alpha)(Z\alpha)^4$, are 
determined by $c_D$.  
Our definition (\ref{eq:re_define}) of the proton radius in the presence of 
radiative corrections implies that 
the $Z^2\alpha (Z\alpha)^4$ correction is unchanged from 
the point-particle result, so that~\cite{HillPaz} 
\begin{multline}\label{eq:vertex}
\delta E^{\rm vertex}(n,\ell) = {2 m_r^3 (Z\alpha)^4 (r_E^p)^2 \over 3 n^3}\delta_{\ell 0} 
\\
+ {m_r^3 Z^2\alpha (Z\alpha)^4 \over \pi n^3} \bigg[ 
\delta_{\ell 0}\left( \frac43 \ln{m_p\over m_r\alpha^2} + {10\over 9} \right) 
- \frac43\ln k_0(n,\ell) \bigg] \,.
\end{multline}
Two-photon exchange corrections, of order $(Z\alpha)^5$, are determined by $d_2$.
Considering (\ref{eq:disp}), it is natural to decompose the correction as 
\be
\delta E^{{\rm two}-\gamma}= \delta E^{\rm proton}+\delta E^{\rm continuum}+\delta E^{W_1(0,Q^2)}.
\ee
It is convenient to subtract $\lim_{Q^2\to 0}W_1(0,Q^2)$ from $W_1(0,Q^2)$ in (\ref{eq:disp}), 
and add $\lim_{Q^2\to 0}W_1(0,Q^2)$  
to $W_1^{p,1}(\nu,Q^2)$. 
Infrared singular terms in (\ref{eq:threshold}) 
are then confined to the proton pole contribution.  
Having fixed this terminology, 
we proceed to discuss each of the three terms in $\delta E^{{\rm two}-\gamma}$ in turn.
Our discussion so far applies to general hydrogenic bound states.   
To investigate numerical results we now specialize to muonic hydrogen (``$\mu H$''). 

\noindent
\emph{Proton pole contribution.}
We content ourselves with a simple dipole model for the elastic form factors, 
\begin{align}\label{eq:dipole}
G_E(q^2) \approx {G_M(q^2)/G_M(0)} \approx [1 - q^2/\Lambda^2]^{-2} \,,
\end{align}
where $G_E \equiv F_1+ (q^2/4 m_p^2) F_2$, $G_M \equiv F_1 + F_2$ 
and $\Lambda^2= 0.71\,{\rm GeV}^2$. 
We return to a more sophisticated analysis of this contribution, 
and analogous spin-dependent contributions, 
in forthcoming work~\cite{HillPaz}. 
After isolating the finite term in (\ref{eq:threshold}),
for muonic hydrogen
\begin{align}\label{eq:proton}
\delta E^{\rm proton}_{\mu H}(nS) &\approx (8 / n^3) \left(  0.016 \, {\rm meV} \right) \,.
\end{align}
We refrain from giving a detailed error estimation here; for
the purpose of explaining the muonic hydrogen anomaly, an
error smaller than $100\,\%$ does not have very substantial impact.

\noindent
\emph{Continuum contribution.}
A recent determination of the continuum contribution is~\cite{Carlson:2011zd}
\begin{align}
\delta E^{\rm continuum}_{\mu H}(nS) &\approx (8/ n^3) \left( - 0.0127(5) \, {\rm meV} \right) \,,
\end{align} 
in line with previous results, $-0.014(2)\,{\rm meV}$~\cite{Pachucki:1999zza}, 
$-0.016(3)\,{\rm meV}$~\cite{Martynenko:2005rc}. 

\noindent
\emph{$W_1(0,Q^2)$ contribution.}
In the SIFF model (\ref{eq:SIFF}) one finds, 
\be\label{eq:SIFFresult}
\delta E^{W_1(0,Q^2),{\rm SIFF}}_{\mu H}(nS) = (8/ n^3) \left( - 0.034 \, {\rm meV} \right)  \,. 
\ee 
The sum of the proton pole and $W_1(0,Q^2)$ contributions 
in this model, $0.016\,{\rm meV} -0.034\,{\rm meV} =-0.018\,{\rm meV}$, 
reproduces previous results~\cite{Pachucki:1996zza}. 
It is not hard to construct model functions for $W_1(0,Q^2)$ 
that have the correct small-$Q^2$ and 
large-$Q^2$ behavior, but give a much larger contribution than the SIFF model.

\noindent
\emph{Comparison to previous results.}
In order to make the comparison to the literature clearer, we
collect the results of this analysis in Table~\ref{tab}, which
compares numerical results for $\order(\alpha^5)$ proton structure corrections 
in the $2P-2S$ Lamb shift of muonic hydrogen.  
We focus on the two reference sources that were
used in \cite{Pohl:2010zz}, namely \cite{Pachucki:1999zza} and \cite{Borie}. 
These works model $\delta E^{W_1(0,Q^2)}_{\mu H}$ as
a sum of proton and non-proton contributions, 
adding the respective terms to $\delta E^{\rm proton}_{\mu H}$ and $\delta E^{\rm continuum}_{\mu H}$.  
In order to simplify the comparison we present in the table
the total contribution to  $\delta E^{{\rm two}-\gamma}_{\mu H}$  from
\cite{Pachucki:1999zza} and  \cite{Borie}. 
In particular for \cite{Pachucki:1999zza} we add the
$(Z\alpha)^5$ nuclear size correction 
($0.0232$ meV) and the proton polarizability correction ($0.012$ meV). 
For \cite{Borie} we add the $(Z\alpha)^5$ nuclear size correction 
($0.0232$ meV), 
the polarizability correction ($0.015$ meV), 
and the recoil finite size correction ($0.013$ meV). 
In \cite{Pohl:2010zz} the nuclear size correction at order 
$(Z\alpha)^5$ from \cite{Pachucki:1999zza} and \cite{Borie} employs the SIFF ansatz 
(\ref{eq:SIFF}) for $W_1(0,Q^2)$; the $(r_E^p)^3$ scaling employed in \cite{Pohl:2010zz} 
assumes the large $m_p$ limit and a one-parameter model for
$G_E$ and $G_M$.

\begin{table}
\begin{tabular}{ll|c|c|c}
Contribution&&Ref. \cite{Pachucki:1999zza}&Ref. \cite{Borie}&This work\\\hline\hline
$\delta E^{\rm vertex}$ & & $-0.0099$ & $-0.0096$ & $-0.0108$ \\ 
\hline
\multicolumn{1}{l|} {}&$\delta E^{\rm proton}_{\mu H}$&&&$-0.016$\\\cline{2-2}\cline{5-5}
\multicolumn{1}{l|} {$\delta E^{{\rm two}-\gamma}$}& $\delta E^{W_1(0,Q^2)}_{\mu H}$ & $0.035$ & $0.051$ & Model Dependent \\\cline{2-2}\cline{5-5}
\multicolumn{1}{c|} {}&$\delta E^{\rm continuum}_{\mu H}$ && $$ & $0.013 $ \cite {Carlson:2011zd}\\\hline\hline
Total&& $0.025$ & $0.042$ & 
\end{tabular} 
\caption{\label{tab} Comparison between this and previous works for 
 $\order(\alpha^5)$ 
proton structure corrections to the $2P-2S$ Lamb shift in muonic hydrogen, in meV. }
\end{table}

Let us note three differences between our results and 
the theoretical predictions used in \cite{Pohl:2010zz}, and collected in Table~\ref{tab}.
First, 
the $\alpha^5$ proton vertex correction from \cite{Pachucki:1996zza,Pachucki:1999zza}
uses a different convention for the charge radius~\cite{footnoter},
while the result from \cite{Eides:2000xc}, adopted in \cite{Borie}, 
uses a model-dependent SIFF prescription for 
the proton vertex correction; the complete result with the charge 
radius definition (\ref{eq:re_define}) is given by (\ref{eq:vertex}), displayed in the first line of the table.
Second, the ``recoil finite size" of \cite{Borie}, adopted from \cite{Friar:1978wv}, 
is in fact part
of $\delta E^{{\rm two}-\gamma}_{\mu H}$. 
Including it as separate contribution would lead to double counting. 
Third,  the $\delta E^{W_1(0,Q^2)}_{\mu H}$ contribution is model-dependent; 
the current theoretical prediction is based on the SIFF ansatz.  
We conclude that the dominant radiative correction to 
proton structure is subject to uncertainties from unreliable hadronic models. 

\noindent
\textbf{Discussion.}
We have presented the NRQED formalism for systematically analyzing  
proton structure effects in hydrogenic bound states. 
The Lamb shift in muonic hydrogen is sensitive to a new structure-dependent 
contact interaction (\ref{eq:contact}).  
The strength of this interaction is not determined by 
measured proton form factors or inelastic structure functions. 
Taking all other contributions as fixed, the muonic hydrogen Lamb shift determines 
$d_2$ in (\ref{eq:contact}).  NRQED then predicts 
a universal shift for other spin-independent energy splittings in 
muonic hydrogen.

The strength of the contact interaction can be related to a so-far 
poorly constrained piece of the forward Compton amplitude of the proton, 
$W_1(0,Q^2)$. 
In this Letter, we have established some model-independent properties
of $W_1(0,Q^2)$.  Firstly, the $\order(Q^2)$ Taylor expansion (\ref{eq:lowQ2}) 
is determined by NRQED in terms of measured quantities; 
secondly, the asymptotic behavior is determined by OPE techniques 
to be $\sim Q^{-2}$.    
The intermediate region remains poorly constrained~\cite{footnoteres}.
The lack of theoretical control over $W_1(0,Q^2)$ 
introduces theoretical uncertainties that have not been taken into account
in the literature.  
A common approach is to use the SIFF model (\ref{eq:SIFF}), 
but this is not derived from first principles and 
gives the misleading impression that the dominant $Q^2$ dependence 
is constrained by onshell 
form factors~\cite{Pachucki:1999zza,Carlson:2011zd}. 
Such extrapolations represent models for $W_1(0,Q^2)$, typically 
without the correct large $Q^2$ behavior. 
While we do not attempt an explicit modeling of $W_1(0,Q^2)$,  
we believe that the uncertainty assigned to this
contribution ($\lesssim 0.004\,{\rm meV}$~\cite{Pohl:2010zz})
is underestimated by at least an order of magnitude.  

As further applications, the nonrelativistic effective theory for vector fields can 
similarly be employed to describe deuterium.  
The NRQED lagrangian at order $1/M^4$ can be used to systematically 
analyze $\delta E^{{\rm two}-\gamma}$ in the small-lepton mass limit relevant to electronic hydrogen, 
and describes spin polarizabilities of the proton~\cite{HillPaz}.  

\vskip 0.1in
\noindent
\emph{Acknowledgements.}
We thank T.~Becher, S.~Brodsky and C.~Wagner for discussions. 
Work supported by NSF Grant 0855039 and DOE grant DE-FG02-90ER40560.

\end{document}